\documentclass[aps,prd,twocolumn,showpacs,preprintnumbers,superscriptaddress,amsmath,amssymb]{revtex4}

\usepackage{booktabs}
\usepackage{graphicx}
\usepackage{dcolumn}
\usepackage{bm,epsfig}
\usepackage{epstopdf}
\usepackage[normalem]{ulem} 
\usepackage[usenames]{color}
\renewcommand\sout{\bgroup \color{red} \ULdepth=-.5ex \ULset}

\begin{document}

\title{ Possible $S$-wave $ND^{(*)}$ and $N\bar B^{(*)}$ bound states in a chiral quark model}

\author{Dan Zhang}
\email{zhangd@imu.edu.cn}
\affiliation{School of Physical Science and Technology, Inner Mongolia University, Hohhot 010021, People's Republic of China}
\affiliation{Department of Physics and Astronomy, University of Georgia, Athens, Georgia 30602, USA}
\author{Dan Yang}
\email[Corresponding author. Email: ]{942226040@qq.com}
\affiliation{School of Physical Science and Technology, Inner Mongolia University, Hohhot 010021, People's Republic of China}
\author{Xiao-Fei Wang}
\email[Corresponding author. Email: ]{451717166@qq.com}
\affiliation{School of Physical Science and Technology, Inner Mongolia University, Hohhot 010021, People's Republic of China}
\author{Kanzo Nakayama}
\email[Corresponding author. Email: ]{nakayama@uga.edu}
\affiliation{Department of Physics and Astronomy, University of Georgia, Athens, Georgia 30602, USA}

\begin{abstract}
$S$-wave bound-states composed of a nucleon($N$) and a heavy meson ($D$, $D^*$, $\bar{B}$ or $\bar{B}^*$) are investigated in both the chiral SU(3) quark model and the extended chiral SU(3) quark model by solving the resonating group
method equation. The results reveal that the
$ND$ and $ND^*$ interactions in the corresponding relative $S$-wave states are attractive, arising mainly from one boson exchange processes between light quarks. It is shown that these attractions are strong enough to form six $ND$ and $ND^*$ $S$-wave bound states in the extended chiral SU(3) quark model with the binding energies in the range of $3-45$ MeV, and three $S$-wave bound states within the chiral SU(3) quark model with binding energies of $2-8$ MeV.
In particular, the experimentally observed $\Sigma_c(2800)$
is interpreted to be most likely an $S-$wave $ND$ state with the total isopsin $I=0$ and spin-parity $J^P=1/2^-$, while $\Lambda_c(2940)^+$ as an $S-$wave $ND^*$ state with $I=0$ and $J^P=3/2^-$. Further information on the $ND$ and $ND^*$ interactions in the (unbound) scattering kinematics are obtained from the corresponding $S$-wave phase shifts. The $N\bar{B}$ and $N\bar{B}^*$ systems are also investigated within the present two models and some $S-$wave bound states with binding energies in the range of $1-60$ MeV are predicted in these systems: six (in total) within the extended chiral SU(3) quark model and, four, within the chiral SU(3) quark model.
\end{abstract}


\maketitle

\section{Introduction}

In the past decade, many new charmed hadrons have been detected
experimentally (see the review literature \cite{guo-18} and references therein).  Of particular interest among these hadronic states is the two charmed baryonic states: the $\Sigma_c(2800)$ state observed by the Belle Collaboration \cite{belle-05} and the $\Lambda_c(2940)^+$ state reported by the Babar Collaboration \cite{babar-07}. In 2005,
an isospin triplet charmed baryon $\Sigma_c(2800)$ decaying into $\Lambda_c^+ \pi$ was first reported by the Belle Collaboration \cite{belle-05},
who tentatively assigned the quantum numbers $J^P=3/2^-$.
Later, the Babar Collaboration \cite{babar-08} proposed a possible confirmation of the neutral state $\Sigma_c(2800)^0$ with a weak evidence of $J=1/2$, although their
measured mass of $2846\pm 8\pm 10$ MeV is higher than the value quoted by Belle \cite{belle-05}.
In addition, the Babar Collaboration \cite{babar-07} observed a new charmed baryon,
$\Lambda_c(2940)^+$, with a mass of $2939.8\pm 1.3\pm 1.0$ MeV, which was
confirmed by the Belle Collaboration \cite{belle-07} as a resonant structure in the
$\Sigma_c(2455) \pi$ decay with measured mass of $2938.0\pm1.3_{-4.0}^{+2.0}$ MeV.

Since both the $\Sigma_c(2800)$ and $\Lambda_c(2940)^+$ states are just below the
$ND$ and $ND^*$ thresholds, respectively, they are likely explained as the $ND$ and $ND^*$ molecular states \cite{guo-18}, respectively.
Dong \emph{et al.}\cite{ybd-10} considered the isotriplet $\Sigma_c(28 00)$ as a
hadronic molecule composed of a nucleon and a $D$ meson. Their widths of the strong two-body
decay $\Sigma_c \rightarrow \Lambda_c \pi$ for the spin-parity $J^P=3/2^-$ and $J^P=1/2^+$
assignments are consistent with the current data.
In Ref.\cite{cej-09}, based on a coupled channels unitary approach, it is suggested the $\Sigma_c(2800)$ is a
dynamically generated resonance with a dominant $ND$ configuration
having $J^P=1/2^-$. The author of Ref.\cite{jrz-14} proposed
 $\Sigma_c(2800)$ to be an $S$-wave $ND$ state with $J^P=1/2^-$ within the
framework of QCD sum rules. Although its calculated mass is
somewhat larger than the corresponding experimental value, the possibility of
$\Sigma_c(2800)$ to be a molecular state can not be arbitrarily
excluded. Wang \emph{et al.}\cite{zyw-18} assumed that the observed $\Sigma_c(2800)^0$
is an $S-$wave $ND$ molecular state with $J^P=1/2^-$ in the Bethe-Salpeter equation approach.
Concerning the $\Lambda_c(2940)^+$ state, Zhang\cite{jrz-14} suggested it to be an $S$-wave $ND^*$ state with $J^P=3/2^-$ in the
framework of QCD sum rules. The $S$-wave $pD^{*0}$
molecular state with $J^P=1/2^-$ was suggested in Ref.\cite{xgh-07}.
Dong \emph{et al.}\cite{ybd-10-2,ybd-10-3} studied $\Lambda_c(2940)^+$, suggesting it to be an
$ND^*$ molecular state with $J^P=1/2^{\pm}$; their results also
suggest that the spin-parity $J^P=1/2^-$ should be ruled out. He \emph{et al.}\cite{jh-10}
indicated the existence of the $ND^*$ systems with
$J^P=1/2^{\pm},3/2^{\pm}$ which, not only provided valuable
information to underlying the structure of $\Lambda_c(2940)^+$, but also
improves our knowledge of the interaction of nucleon and $D^*$.
A possible molecular candidate for the $\Lambda_c(2940)^+$ with $J^P=3/2^-$ was obtained in a chiral constituent quark model\cite{pgo-13,pgo-16}.

On the other hand, there is an alternative way of theoretical study to consider $\Sigma_c(2800)$ and $\Lambda_c(2940)^+$ as conventional charmed baryons\cite{sca-86,deb-08,hga-07,bch-09,jhe-11}.
In a relativized potential model, the masses of $\Sigma_c(2800)$ and $\Lambda_c(2940)^+$ were close to theoretical values
of $\Sigma_c^*$ with $J^P=3/2^-$ or $J^P=5/2^-$ and $\Lambda_c^*$ with $J^P=5/2^-$ or $J^P=3/2^+$, respectively\cite{sca-86}.
In the relativistic quark-diquark picture, the $\Sigma_c(2800)$ state has been suggested as one of the orbital excitations($1P$) of the ground state $\Sigma_c$ with
$J^P=1/2^-$, $3/2^-$ or $5/2^-$. The $\Lambda_c(2940)^+$ has been proposed as the first radial excitation of $\Sigma_c$ with
$J^P=3/2^+$ \cite{deb-08}. In Ref.\cite{hga-07}, based on the Faddeev method, it is indicated that the $\Sigma_c(2800)$ state would correspond to an orbital excitation with
$J^P=1/2^-$ or $3/2^-$ and that the $\Lambda_c(2940)^+$ state may constitute the second orbital excitation of $\Lambda_c$.
In a mass loaded flux tube model, Chen \emph{et al.} \cite{bch-09} suggested that $\Lambda_c(2940)^+$ could be the orbitally excited $\Lambda_c^+$ with
$J^P=5/2^-$. He \emph{et al.} \cite{jhe-11} evaluated the production rate of $\Lambda_c(2940)^+$ as a charmed baryon in view of future experiments at PANDA.

Despite considerable efforts spent in the study of the $\Sigma_c(2800)$ and $\Lambda_c(2940)^+$ states, they
are not yet fully understood. Thus, it is timely to make further efforts to study these states to reveal their properties.

Earlier investigations indicate that the chiral SU(3) quark model
\cite{xcqm-zhang} and the extended chiral SU(3) quark model
\cite{lrdai03} are successful in studying hadronic systems with
light flavors, such as $NN$, $NY$\cite{xcqm-zhang, lrdai03}, $NK$ \cite{hfnk04, hfnk05}, $\Delta K$, $\Sigma K$\cite{hfdk05}, $N\bar{\Omega}$\cite{zd} interactions,
and the structures of pentaquark states \cite{si2,wu}. Recently, these two quark
models were applied to the heavy flavor sectors and valuable results
were obtained, which include the masses of the singly heavy
ground-state baryons \cite{zhqy}, the tetraquark states including heavy quarks
\cite{hx07,hx08,zm08}, interactions of $DK$ \cite{ls}, $D\bar{D}$
and $B\bar B$ \cite{L}, $\Sigma_c\bar{D}$ and $\Lambda_c\bar{D}$
\cite{sgla}, and structures of $X(3872)$ \cite{x}, $Z_b(10610)$ and
$Z_b(10650)$ \cite{z}.
In this paper, we shall extend the application of these two successful models \cite{xcqm-zhang,lrdai03,hfdk05,zd,si2,wu} to explore the $S$-wave $ND$ and $ND^*$ as well as $N\bar{B}$ and $N\bar{B}^*$ heavy quark systems with all possible quantum numbers. We solve the resonating group method (RGM) equation to obtain the pertinent solutions. In particular, we give an interpretation of the structures of both the $\Sigma_c(2800)$ and $\Lambda_c(2940)^+$ states within these models.
The present models are also applied to the $S$-wave $N\bar B$ and $N\bar B^*$ systems to predict some new bound states.
In addition, some useful information on the $ND^{(*)}$ and $N \bar B^{(*)}$ interactions will be obtained from the corresponding scattering processes.
Throughout this paper, we use the notation $ND^{(*)}$ to indicate both $ND$ and $ND^*$ systems. Likewise, $N\bar{B}^{(*)}$  indicates both $N\bar{B}$ and $N\bar{B}^*$ systems.

We mention that the current models are different from the chiral constituent quark model of Refs.\cite{pgo-13,pgo-16} in the details of the assumed Hamiltonian.
In particular, we consider the vector-meson exchange contributions explicitly (extended chiral SU(3) quark model), while this is absent in Refs.\cite{pgo-13,pgo-16}.

The paper is organized as follows: Sec. II presents a brief description of the chiral
and extended chiral SU(3) quark models and associated parameter values used in the present work, as well as the resonating group method. In Sec. III,
we discuss the numerical results. Finally, Sec. IV gives a summary and conclusion.

\section{Formalism} \label{sec:Formalism}
\subsection{Model}

The details of the chiral and extended chiral SU(3) quark
models considered in the present work have been described in Refs.~\cite{xcqm-zhang, lrdai03}. Thus, here,
we only summarize the most relevant aspects of the model to the present work. The Hamiltonian of the
baryon($qqq$)-meson($Q\bar q$) ($q$ stands for light quark and, $Q$, for heavy
quark) system can be written as
\begin{eqnarray}
H=\sum_{i}T_i-T_{cm}+\sum_{i<j} V_{ij}\; ,
\end{eqnarray}
where $T_i$ is the kinetic energy operator for a single quark specified by the subscript $i$ and, the summation runs over the quarks in the baryon-meson system under consideration.
$T_{cm}$ is the kinetic energy operator associated with the center-of-mass (c.m.) motion.
Note that the latter is subtracted from the first term in the above equation such that the Hamiltonian accounts for the kinetic energy associated with the relative motion of the quarks.
$V_{ij}$ represents the interaction between two
quarks specified by the subscripts $i$ and $j$, and it includes the interactions between $qq$ inside the baryon($V_{qq}$),
$Q\bar q$ inside the meson($V_{Q\bar q}$), and $Qq$ or $q\bar q$ between the baryon and
the meson($V_{Qq}$ or $V_{q\bar q})$.

For a given $qq$ pair, there are three parts in the corresponding interaction, i.e.,
\begin{equation}
V_{qq}(ij)=V^{conf}(ij)+V^{OGE}(ij)+V^{ch}(ij).
\label{eq:Vqq}
\end{equation}

Below, we specify each of the terms in the above equation.
First, $V^{OGE}$ stands for the one-gluon-exchange (OGE) interaction,
\begin{eqnarray}
\label{oge}
V^{OGE}_{qq}(ij)&=&\frac{1}{4}g_{i}g_{j}\left(\lambda^c_i\cdot\lambda^c_j\right)\left\{\frac{1}{r_{ij}}-\frac{\pi}{2} \delta({\bm r}_{ij})\left(\frac{1}{m^2_{i}}\right.\right. \nonumber\\
 & &\left.\left.+\frac{1}{m^2_{j}}+\frac{4}{3}\frac{1}{m_{i}m_{j}}
({\bm \sigma}_i \cdot {\bm \sigma}_j)\right)\right\},
\label{eq:Voge}
\end{eqnarray}
where ${\bm r}_{ij}$ is the relative coordinate of two quarks specified by the subscripts $i$ and $j$,
$g_i(g_j)$ denotes the OGE coupling constant,
and $m_i(m_j)$ is the mass of the $i$th ($j$th) quark.
$(\lambda^c_i \cdot \lambda^c_j)$ and $({\bm \sigma}_i \cdot {\bm \sigma}_j)$ are the operators in the color and the spin spaces, respectively.

$V^{ch}$ in Eq.~(\ref{eq:Vqq}) represents the interaction from chiral fields coupling, which includes the scalar and
pseudoscalar meson exchanges in the chiral SU(3) quark model,
\begin{eqnarray}
V^{ch}_{qq}(ij) = \sum_{a=0}^8 V_{\sigma_a}({\bm r}_{ij})+\sum_{a=0}^8
V_{\pi_a}({\bm r}_{ij}) ,
\end{eqnarray}
where $\sigma_a$($\pi_a$) $(a=0, \cdots, 8)$ denotes the low-lying scalar-(pseudoscalar-)meson nonet fields.
In the above equation,
\begin{eqnarray}
V_{\sigma_a}({\bm r}_{ij})&=&-C(g_{ch},m_{\sigma_a},\Lambda)
X_1(m_{\sigma_a},\Lambda,r_{ij})\nonumber\\
& & \times [\lambda_a(i)\lambda_a(j)],
\end{eqnarray}
and
\begin{eqnarray}
V_{\pi_a}({\bm r}_{ij})&=&C(g_{ch},m_{\pi_a},\Lambda)
\frac{m^2_{\pi_a}}{12m_{i}m_{j}} X_2(m_{\pi_a},\Lambda,r_{ij})\nonumber\\
&&\times ({\bm \sigma}_i\cdot{\bm \sigma}_j) [\lambda_a(i)\lambda_a(j)],
\end{eqnarray}
where
\begin{eqnarray}
C(g_{ch},m,\Lambda)=\frac{g^2_{ch}}{4\pi}
\frac{\Lambda^2}{\Lambda^2-m^2} m,
\end{eqnarray}
\begin{eqnarray}
\label{x1mlr} X_1(m,\Lambda,r)=Y(mr)-\frac{\Lambda}{m} Y(\Lambda r) ,
\end{eqnarray}
\begin{eqnarray}
X_2(m,\Lambda,r)=Y(mr)-\left(\frac{\Lambda}{m}\right)^3 Y(\Lambda
r),
\end{eqnarray}
\begin{eqnarray}
Y(x)=\frac{1}{x}e^{-x},
\end{eqnarray}
with $m_{\sigma_a}$($m_{\pi_a}$) denoting the mass of the scalar(pseudoscalar) meson,
and $\Lambda$ the cutoff mass for mesons. $g_{ch}$ represents the coupling constant for the
scalar and pseudoscalar chiral field couplings. $[\lambda_a(i)\lambda_a(j)]$ is the operator in the flavor space.
Note that the flavor SU(3) classification of the low-lying scalar mesons is not well established, since the underlying structures of these mesons are still an open issue.
Nevertheless, considering these mesons as members of the SU(3) nonet seems to work rather well \cite{xcqm-zhang,lrdai03,hfnk04,hfnk05,zd,si2,wu,zhqy,hx07,hx08,zm08,ls,L,sgla,z}.

The extended chiral SU(3) quark model includes the vector meson exchanges in addition, viz.,
\begin{eqnarray}
V^{ch}_{qq}(ij) = \sum_{a=0}^8 V_{\sigma_a}({\bm r}_{ij})+\sum_{a=0}^8
V_{\pi_a}({\bm r}_{ij})+\sum_{a=0}^8 V_{\rho_a}({\bm r}_{ij}).
\end{eqnarray}
where $\rho_{a}(a=0,..,8)$ denotes the low-lying vector-meson nonet fields. In the above equation,
\begin{eqnarray}
V_{\rho_a}({\bm r}_{ij})&=&C(g_{chv},m_{\rho_a},\Lambda)\left\{
X_1(m_{\rho_a},\Lambda,r_{ij})\right.\nonumber\\
&&\left.\left.+ \frac{m^2_{\rho_a}}{6m_{i}m_{j}}
\left(1+\frac{f_{chv}}{g_{chv}}\frac{m_{i}+m_{j}}{M_P}\right.\right.\right.\nonumber\\
&&\left.+\frac{f^2_{chv}}{g^2_{chv}}\frac{m_{i}m_{j}}{M^2_P}\right)
X_2(m_{\rho_a},\Lambda,r_{ij}) \nonumber \\
&& \left. \cdot ({\bm \sigma}_i\cdot{\bm \sigma}_j)\right\}[\lambda_a(i)\lambda_a(j)],
\end{eqnarray}
with $m_{\rho_a}$ denoting the mass of the vector meson and $M_P$ is a mass scale, which is taken to be the proton mass.
$g_{chv}$ and $f_{chv}$ are the coupling
constants associated with the vector and tensor couplings of the vector
meson fields, respectively.

The flavor singlet-octet mixing ($\pi_0$-$\pi_8$) of the pseudoscalar mesons is accounted for according to
\begin{eqnarray}
\eta=\pi_8 \textmd{cos}\theta^{ps}-\pi_0\textmd{sin}\theta^{ps},\nonumber\\
\eta'=\pi_8 \textmd{sin}\theta^{ps}+\pi_0\textmd{cos}\theta^{ps},
\label{eq:psSOMix}
\end{eqnarray}
with the mixing angle of $\theta^{ps}=-23^{\circ}$ \cite{lrdai03,hfnk04,hfnk05,wu,ls}.
Analogously, the mixing angle $\theta^s$ corresponding to the singlet-octet ($\sigma_0$-$\sigma_8$) mixing of the scalar mesons
is taken to be $\theta^s=0^{\circ}$ \cite{xcqm-zhang, lrdai03, zd,ls}, i.e., the $\sigma$ meson is assumed to be a pure singlet ($\sigma_0$) and $f_0$, a pure octet ($\sigma_8$) meson, respectively.
Since the vector mesons $\omega$ and $\phi$ are nearly ideally mixed states of $\rho_0$ and $\rho_8$, they are approximated to be pure $(u\bar u+d\bar d)$ and $s\bar s$ states, respectively \cite{hfnk04,hfnk05,zd}, i.e., $\theta^v \approx \theta^v_{ideal} =  -54.736^{\circ}$.

$V^{conf}$ in Eq.~(\ref{eq:Vqq}) stands for the confinement potential, taken as the linear form in this
work following Refs.~\cite{zhqy,hx07,hx08,zm08,ls,L,sgla},
\begin{eqnarray}
\label{conf}
V^{conf}_{qq}(ij)=-(\lambda_{i}^{c}\cdot\lambda_{j}^{c})(a_{ij}r_{ij}+a_{ij}^{0}),
\label{eq:Vconf}
\end{eqnarray}
with $a_{ij}$ denoting the confinement strength and $a^{0}_{ij}$
the zero-point energy.

The quark-anti-quark interaction $V_{q \bar q}$ can be obtained from the quark-quark interaction $V_{qq}$ specified above by simple transformations \cite{hfnk04,hfnk05,hfdk05,zd,si2,wu}. For $V^{OGE}_{q \bar q}$ and $V^{conf}_{q\bar q}$, the
transformation is given by the replacement $(\lambda _i^c\cdot \lambda
_j^c)\rightarrow (-\lambda _i^c\cdot \lambda _j^{c*})$ in $V^{OGE}_{q q}$ and $V^{conf}_{q q}$ given by Eq.~(\ref{eq:Voge}) and (\ref{eq:Vconf}), respectively,
while for $V_{q\bar{q}}^{ch}$, we have
\begin{equation}
V_{q\bar{q}}^{ch}=\sum_{j}(-1)^{G_j}V_{qq}^{ch,j}.
\end{equation}
Here $(-1)^{G_j}$ represents the G parity of the $j$th meson.

In the heavy quark sector, chiral symmetry is explicitly broken and, therefore, $V^{ch}$ is not considered in the interactions involving heavy quarks \cite{pgo-13,zhqy,hx07,hx08,zm08,ls,L,sgla,x,z}. Hence,
\begin{equation}
V_{Qq}(ij)=V^{conf}(ij)+V^{OGE}(ij)  ,
\end{equation}
and likewise for $V_{Q \bar q}$.
The confinement and one-gluon-exchange potentials, $V^{conf}(ij)$ and $V^{OGE}(ij)$, in the above equation can be obtained from those in $V_{qq}$ and
$V_{q\bar q}$, respectively, by replacing the mass of a light quark by that of a heavy quark.

Note that, since we confine ourselves to the $S$ partial waves in this work, the spin-orbit and tensor forces - in principle present in Eq.(3), (5), (6), and  (12)- do not play any role.

\subsection{The framework of resonating group method}

The resonating group method (RGM) is a well
established method for learning about the interaction between two
clusters, which has been widely used in Nuclear Physics and in constituent quark models \cite{pgo-16,xcqm-zhang,lrdai03,hfnk04,hfnk05,hfdk05,zd,si2,wu,ls,L,sgla,x,z,o,p,q}.
For the systems composed of the baryon $N$ and meson $M_Q$, the wave function of the five-quark system is taken as
\begin{eqnarray}
\Psi={\cal{A}} [ \phi_N(\bm \xi_1,\bm \xi_2) \phi_{M_Q}(\bm \xi_3)\chi({\bm
R}_{NM_Q})]_{\alpha},
\label{eq:Psi}
\end{eqnarray}
where ${\bm \xi}_1$, ${\bm \xi}_2$ are the internal coordinates for
the cluster $N$, and ${\bm \xi}_3$ is the internal
coordinate for the cluster $M_Q$,
\begin{eqnarray}
{\bm \xi}_1&=&{\bm r}_2-{\bm r}_1,~~{\bm \xi}_2={\bm r}_3-\frac{m_1{\bm r}_1+m_2{\bm r}_2}{m_1+m_2}\nonumber\\
{\bm \xi}_3&=&{\bm r}_5-{\bm r}_4.
\end{eqnarray}
${\bm R}_{NM_Q}\equiv {\bm R}_N-{\bm R}_{M_Q}$ is the
relative coordinate between the two clusters, $N$ and $M_Q$.
$\alpha$ represents the set of all quantum numbers to specify a state of the baryon-meson system.

$\phi_N(\bm \xi_1,\bm \xi_2)$ and $\phi_{M_Q}(\bm \xi_3)$ in Eq.(\ref{eq:Psi}) are the internal cluster wave functions of $N$ and
$M_Q$, respectively. They are given by

\begin{eqnarray}
\phi_N(\bm \xi_1,\bm \xi_2) & = & \left( \frac{m_{{\bm \xi}_1} \omega}{\pi} \right)^{3/4} \textmd{exp}\left( -\frac{m_{{\bm \xi}_1} \omega}{2} \bm \xi_1^2\right)\nonumber\\
 & & \cdot\left( \frac{m_{{\bm \xi}_2} \omega}{\pi} \right)^{3/4} \textmd{exp}\left( -\frac{m_{{\bm \xi}_2} \omega}{2} \bm \xi_2^2\right),
\label{eq:N}
\end{eqnarray}

\begin{eqnarray}
\phi_{M_Q}(\bm \xi_3)  =  \left( \frac{m_{{\bm \xi}_3} \omega}{\pi} \right)^{3/4} \textmd{exp}\left( -\frac{m_{{\bm \xi}_3} \omega}{2} \bm \xi_3^2\right),
\label{eq:M}
\end{eqnarray}
with
\begin{eqnarray}
m_{{\bm \xi}_1}&=&\frac{m_1m_2}{m_1+m_2},~~m_{{\bm \xi}_2}=\frac{(m_1+m_2)m_3}{m_1+m_2+m_3}\nonumber\\
m_{{\bm \xi}_3}&=&\frac{m_4m_5}{m_4+m_5},
\end{eqnarray}
and
\begin{eqnarray}
\omega=\frac{1}{m_u b_u^2}.
\label{eq:Bu}
\end{eqnarray}

${\cal A}$ in Eq.(\ref{eq:Psi}) is the antisymmetrization operator, defined as
\begin{equation}
{\cal A}\equiv1-\sum_{i=1}^3 P_{i4}= 1-3P_{34},
\end{equation}
where $P_{i4}$ represents the permutation operator between the three quarks in the cluster $N$ and the heavy quark in $M_Q$.
Note that, since there is no quark exchange between the two color-singlet clusters $N$ and $D^{(*)}$ or $\bar B^{(*)}$, the antisymmetrization operator when acting on the quarks between these two clusters becomes ${\cal A}=1$, i.e., $P_{34} = 0$.
The consequence of this is that the matrix elements of the color operator $(\lambda _i^c\cdot \lambda _j^c)$ in Eq.(\ref{eq:Voge}) and Eq.(\ref{eq:Vconf}) vanish identically, which lead to the absence of the one-gluon-exchange and confinement potentials between the clusters $N$ and $D^{(*)}$ or $\bar B^{(*)}$. These potentials, however, act between the quarks within the individual clusters $N$ and $D^{(*)}$ or $\bar B^{(*)}$.
Hence, the OGE and confinement potentials affect the binding energies (cf. Eq.~(\ref{eq:BE})) of the bound $ND^{(*)}$ and $N\bar{B}^{(*)}$ states only indirectly through the calculated masses of the individual clusters $N$, $D^{(*)}$ and $\bar{B}^{(*)}$. Moreover, to the extent that the parameters of the OGE and confining potentials are adjusted to reproduce the masses of these individual clusters as explained later in this section, they have no effect on the calculated binding energies of the $ND^{(*)}$ and $N\bar{B}^{(*)}$ bound states.

Finally,
$\chi({\bm R}_{N M_Q})$ in Eq.(\ref{eq:Psi}) is the relative
wave function of the two clusters, which can be obtained by solving the following equation
\begin{eqnarray}
\int \phi_N^+({\bm \xi}_1, {\bm \xi}_2)\phi_{M_Q}^+({\bm \xi}_3)(H-E)\Psi \textmd{d}{\bm \xi}_1 \textmd{d}{\bm \xi}_2 \textmd{d}{\bm \xi}_3 =0.
\end{eqnarray}

With $\Psi$, $\phi_N(\bm \xi_1,\bm \xi_2)$ and $\phi_{M_Q}(\bm \xi_3)$ given by Eqs. (16,18,19), the RGM equation becomes
\begin{eqnarray}
\int {\cal L}({\bm R}',{\bm R})\chi({\bm R})\textmd{d}{\bm R}=0,
\end{eqnarray}
with
\begin{eqnarray}
{\cal L}({\bm R}',{\bm R})\equiv {\cal H}({\bm R}',{\bm R})-E{\cal N}({\bm R}',{\bm R}),
\end{eqnarray}

\begin{eqnarray}
{\cal H}({\bm R}',{\bm R})&=&\int \phi_N^+ ({\bm \xi}_1, {\bm \xi}_2) \phi_{M_Q}^+({\bm \xi}_3) \delta({\bm R}'-{\bm R}_{NM_Q})\cdot H\nonumber\\
 & \cdot& \phi_N ({\bm \xi}_1, {\bm \xi}_2) \phi_{M_Q}({\bm \xi}_3)\delta({\bm R}-{\bm R}_{NM_Q})  \nonumber\\
&\cdot& \textmd{d}{\bm \xi}_1 \textmd{d}{\bm \xi}_2 \textmd{d}{\bm \xi}_3 \textmd{d}{\bm R}_{NM_Q},
\end{eqnarray}
and
\begin{eqnarray}
{\cal N}({\bm R}',{\bm R})&=&\int \phi_N^+ ({\bm \xi}_1, {\bm \xi}_2) \phi_{M_Q}^+({\bm \xi}_3) \delta({\bm R}'-{\bm R}_{NM_Q})\cdot {\bm 1}\nonumber\\
 & \cdot& \phi_N ({\bm \xi}_1, {\bm \xi}_2) \phi_{M_Q}({\bm \xi}_3)\delta({\bm R}-{\bm R}_{NM_Q})  \nonumber\\
&\cdot& \textmd{d}{\bm \xi}_1 \textmd{d}{\bm \xi}_2 \textmd{d}{\bm \xi}_3 \textmd{d}{\bm R}_{NM_Q}.
\end{eqnarray}

By solving the RGM equation (25), we can obtain the binding energies or scattering phase-shifts and cross sections for the
two-cluster systems. The details of solving the RGM equation can be
found in Refs. \cite{o,p,q,r}.

\subsection{Model parameters}

{\small
\begin{table}[htb]
\caption{\label{para} Model parameters for the light quarks. The
meson masses: $m_{a_0}=980$ MeV,
$m_{f_0}=980$ MeV, $m_{\pi}=138$ MeV, $m_{\eta}=549$ MeV,
$m_{\eta'}=957$ MeV, $m_{\rho}=770$ MeV, $m_{\omega}=782$ MeV,
$m_{\phi}=1020$ MeV.  The cutoff masses: $\Lambda=1100$ MeV for all mesons.
The mixing angles between the flavor singlet and octet mesons: $\theta^{ps}=-23^{\circ}$, $\theta^s=0^{\circ}$, $\theta^v=\theta^v_{ideal}=-54.736^{\circ}$.
}
\begin{center}
\begin{tabular}{cccc}
\hline\hline
  & $\chi$-SU(3)QM & \multicolumn{2}{c}{Ex. $\chi$-SU(3) QM}  \\
  &   I   &    II    &    III \\  \cline{3-4}
  &  & $f_{chv}/g_{chv}=0$ & $f_{chv}/g_{chv}=2/3$ \\
\hline
 $b_u$ (fm)  & 0.5 & 0.45 & 0.45 \\
 $m_u$ (MeV) & 313 & 313 & 313 \\
 $g_u(=g_d)$       & 0.875 & 0.236 & 0.363 \\
 $g_{ch}$    & 2.621 & 2.621 & 2.621  \\
 $g_{chv}$   & $-$   & 2.351 & 1.973  \\
 $m_\sigma$ (MeV) & 595 & 535 & 547 \\
 $a_{uu}$ (MeV/fm) & 87.5 & 75.3 & 66.2 \\
 $a^{0}_{uu}$ (MeV)  & $-$77.4 & $-$99.3 & $-$86.6 \\
\hline\hline
\end{tabular}
\end{center}
\end{table}}

For light quarks, all the model parameters are taken from our previous work
\cite{xcqm-zhang,lrdai03,hfnk04,hfnk05,hfdk05,zd,si2,wu}, which can give a satisfactory description
of the energies of the baryon ground states, the binding energy of
deuteron, and the $NN$ scattering phase-shifts. The harmonic-oscillator
width parameter $b_u$ in Eq.(\ref{eq:Bu}) is taken with different values for the two
models: $b_u=0.50$ fm in the chiral SU(3) quark model and $b_u=0.45$
fm in the extended chiral SU(3) quark model. The up (down) quark mass, $m_u(m_d)$, is taken to be the usual value
of $m_u= m_d =313$ MeV. The coupling constant for the
scalar and pseudoscalar chiral field couplings, $g_{ch}$, is
determined according to the relation
\begin{eqnarray}
\frac{g^{2}_{ch}}{4\pi} = \left( \frac{3}{5} \right)^{2}
\frac{g^{2}_{NN\pi}}{4\pi} \frac{m^{2}_{u}}{M^{2}_{N}},
\end{eqnarray}
with the empirical value $g^{2}_{NN\pi}/4\pi=13.67$\cite{xcqm-zhang}.
The vector-meson field coupling constant $g_{chv}$ in Eq.(12) is taken to be $g_{chv}=2.351$ and $g_{chv}=1.973$ corresponding to the ratio $f_{chv}/g_{chv}=0$ and $f_{chv}/g_{chv}=2/3$, respectively.
The OGE coupling constants, $g_{u}(=g_{d})$ can be determined by the mass splits between
$N$ and $\Delta$. The confinement strengths $a_{uu}$ is fixed by
the stability conditions of $N$, and the zero-point energies $a^{0}_{uu}$
by fitting the masses of $N$.
The masses of the mesons are taken to be those determined
experimentally, except for the $\sigma$ meson, whose value
is adjusted to fit the binding energy of the deuteron. The cutoff
radius $\Lambda^{-1}$ (inverse of the cutoff mass $\Lambda$) is taken to be the value close to the chiral
symmetry breaking scale\cite{ito90,amk91,abu91,emh91}, and common to all mesons.

All the parameter values of the present models associated with the light quarks are displayed in Table \ref{para},
where the first set corresponds to the chiral SU(3) quark model (I), the
second and third sets are for the extended chiral SU(3) quark model
by taking the ratio $f_{chv}/g_{chv}$ to be $0$ (II) and $2/3$ (III),
respectively.

For heavy quarks, to examine their mass dependence of the results, we have first allowed them to take several typical values \cite{zhqy,x}: $m_c=1430$
MeV\cite{hx07,hx08,zm08,ls,L,sgla,z}, $m_c=1550$ MeV\cite{ls,sgla,jv04}, $m_c=1870$ MeV\cite{ls,sgla,bs93} and,
$m_b=4720$ MeV\cite{L,z}, $m_b=5100$ MeV\cite{jv05}, $m_b=5259$
MeV\cite{bs93}. Our numerical results indicate that the heavy-quark-mass dependence
is small and, therefore, we only report the results corresponding to $m_c=1430$ MeV and $m_b=4720$ MeV, which are the values used in the earlier work\cite{zhqy,x,hx07,hx08,zm08,ls,L,sgla,z}.

The OGE coupling constants $g_{Q}$ and the confinement
strengths $a_{Qq}$, and the zero-point energies $a^{0}_{Qq}$ can be determined by fitting the
masses of the mesons with single heavy quark\cite{zhqy,hx07,hx08,zm08,ls}. Thus, in the present study, following Refs.~\cite{zhqy,zm08},  the OGE coupling constants $g_c$ and $g_b$ are taken to be $g_c=0.58$ and $g_b=0.52$, respectively.
The confinement parameters $a_{Qq}$ and $a^{0}_{Qq}$ are adjusted such that the calculated heavy-meson masses be close to the
corresponding experimental values.
The resulting model parameters for the $c$ quark, $a_{cu}$ and $a_{cu}^0$ (together with the OGE coupling constant $g_c$), in the chiral SU(3)
quark model(set I) with $m_c=1430$ MeV are displayed in Table \ref{parah}.
The three sets of adjusted values of  $a_{cu}$ and $a_{cu}^0$ shown there illustrate the sensitivity of the $D$ and $D^*$ meson masses to these parameter values. Similar results/sensitivity are found for the corresponding parameters in the extended chiral SU(3) quark model.  We refrain from showing them here. The above findings concerning the confining potential involving the parameters $a_{bu}$ and $a_{bu}^0$ hold for the case of $b$ quarks as well.  As pointed out before, although the OGE and confining potentials enter in the calculations of the individual clusters $N$, $D^{(*)}$ and $\bar{B}^{(*)}$, they do not play any role in the interaction between the clusters $N$ and $D^{(*)}$ or $\bar{B}^{(*)}$, since these potential contributions vanish identically.
This means that the interaction between $N$ and $D^{(*)}$ or $\bar{B}^{(*)}$ is not sensitive to the light-heavy-quark  parameters $g_Q, a_{Qq}$, and $a^0_{Qq}$; therefore, we only discuss the results of one set of them, namely, that of the first row in Table~\ref{parah}.

{\small
\begin{table}[htb]
\caption{\label{parah} Model parameters for the $c$ quark in the chiral SU(3) quark model(set I) with $m_c=1430$ MeV.
The corresponding calculated and experimental\cite{pdg} values of the masses of the $D$ and $D^*$ mesons are listed as well.}
\begin{center}
\begin{tabular}{ccccc}
\hline\hline
$g_c$& $a_{cu}$ (MeV/fm)&$a_{cu}^0$(MeV)&$D$(MeV)&$D^*$(MeV)\\
\hline
 0.58  & 275.6 & -169.0 & 1869.6&1974.8 \\
       & 275.5 & -162.9 & 1901.8&2007.0 \\
       & 275.2 & -165.0 & 1889.6&1994.8  \\
\hline
Exp. data\cite{pdg}& &  & 1869.6&2006.9 \\
\hline\hline
\end{tabular}
\end{center}
\end{table}}

\section{Results and discussions} \label{sec:results}

\begin{figure*}[htb]
\centerline{
\includegraphics[width=6.0cm,height=10.5cm]{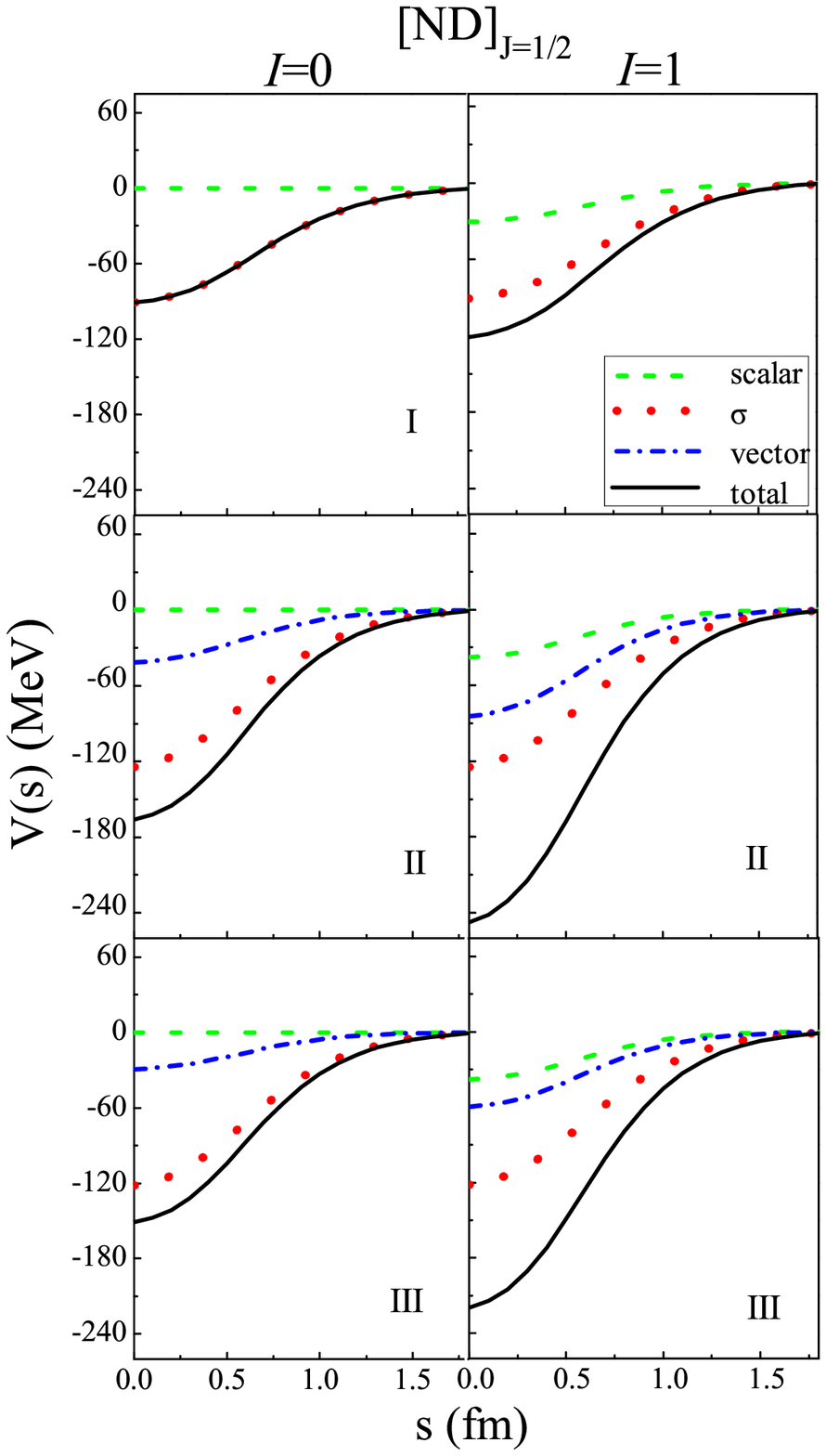}
\includegraphics[width=6.0cm,height=10.4cm]{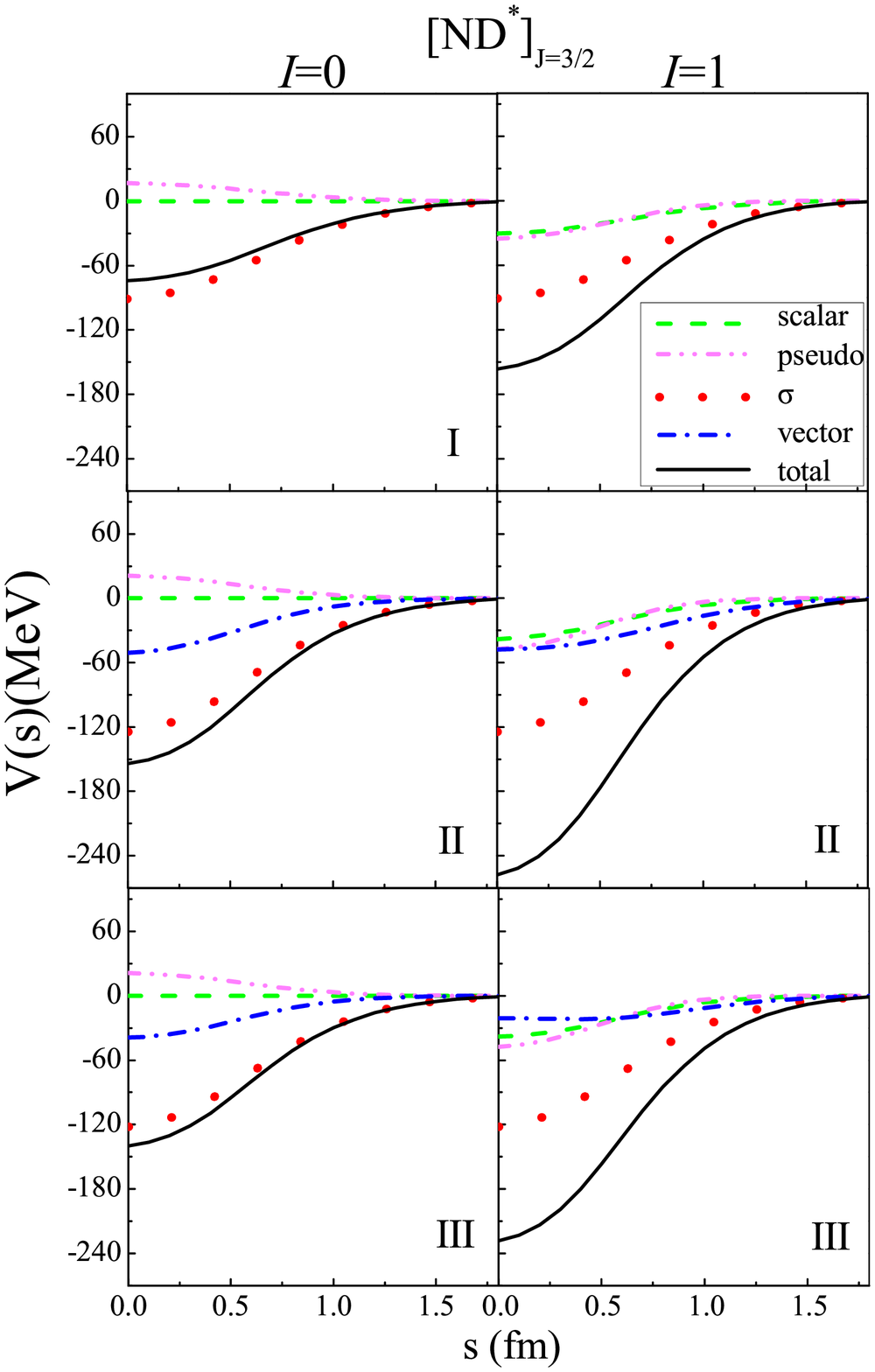}
\includegraphics[width=6.0cm,height=10.4cm]{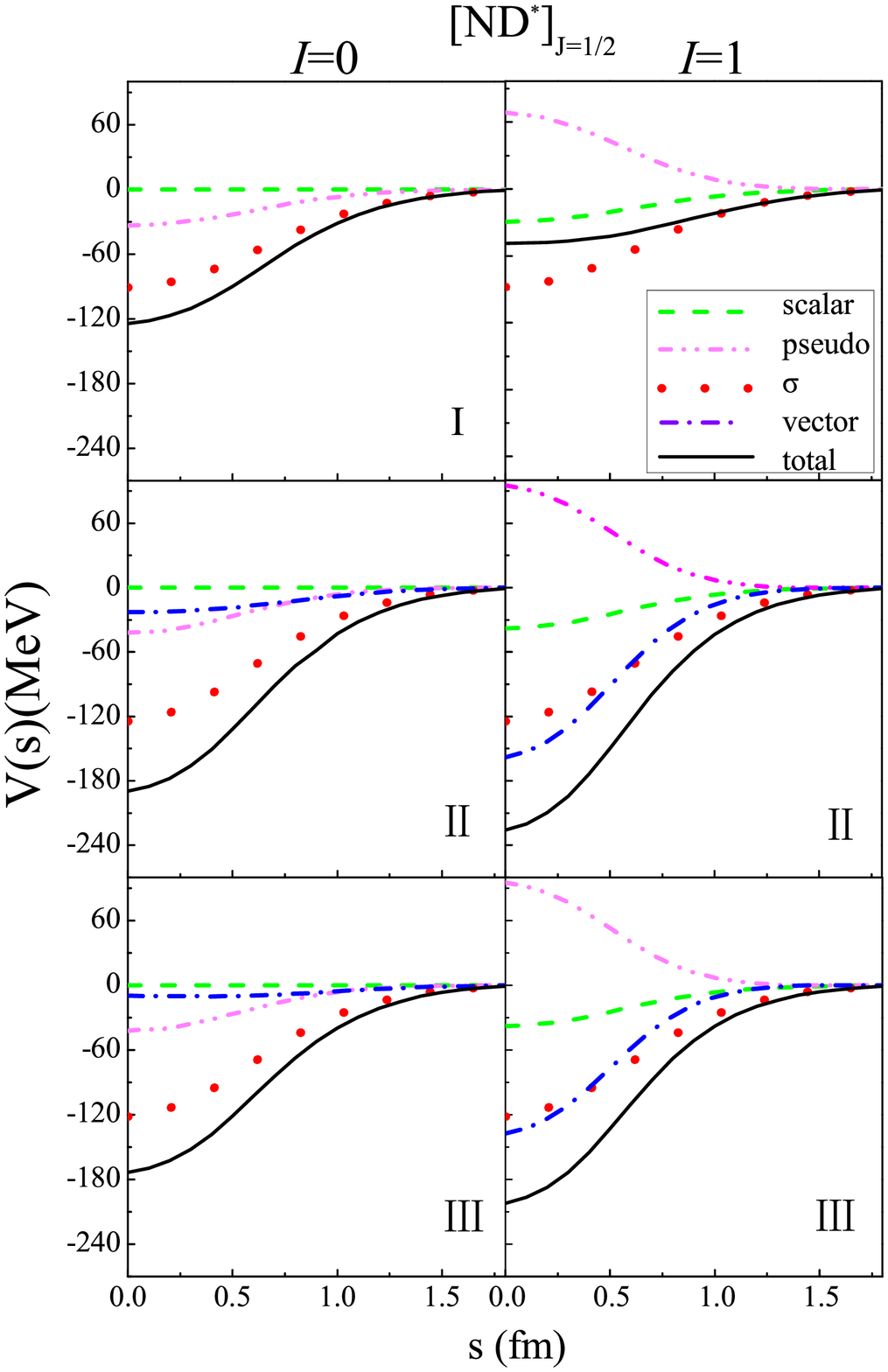}}
\caption{\small \label{vd12}
The effective potentials between the two clusters $N$ and $D$ and, $N$ and $D^*$, corresponding to the $S$-wave $[ND]_{J=1/2}$ (left two columns), $[ND^*]_{J=3/2}$ (middle two columns) and  $[ND^*]_{J=1/2}$ (right two columns)
states as functions of the generator coordinate $s$. The two columns within a given state in the $ND$ or $ND^*$ system as indicated, correspond to the total isospin $I=0$ (left column) and $I=1$ (right column). The panels in the upper rows correspond to the chiral SU(3) quark model (set I). The panels in the middle and lower rows correspond to the extended chiral SU(3) quark model with the parameter sets II and III as given in Table.~\ref{para}, respectively.
The dotted (red) lines represent the effective potentials due to the $\sigma$-meson exchange ($\sigma$).
The dashed (green) lines are the sum of the contributions from the scalar-meson($a_0,f_0$) exchanges (scalar), and the dash-double-dotted (magenta) lines are the sum of the contributions from the pseudoscalar-meson($\pi,\eta, \eta'$) exchanges(pseudo).
The dash-dotted (blue) lines are those from the vector-meson($\omega$, $\rho$) exchanges (vector) in the extended chiral SU(3) quark model(sets II and III).
The solid (black) lines present the sum of all the individual contributions to the  effective potentials (total).}
\end{figure*}

In this work we investigate the $S$-wave $ND^{(*)}$ and $N\bar{B}^{(*)}$ systems within the chiral and extended chiral SU(3) quark models. We consider these systems
in both the bound and unbound (scattering) kinematics.
The spin of $N$ is $S=1/2$, of $D$ and $\bar{B}$ is $S=0$, and of $D^*$ and $\bar{B}^*$ is $S=1$. The isospin of all these five cluster particles is $I=1/2$.
Therefore, the $ND$ and $N\bar{B}$ systems in the $S$ partial-waves can form the states with total spin ($J$) and isospin ($I$), $JI$, with $J=1/2$ and $I=0,1$. Likewise, the $ND^*$ and $N\bar{B}^*$ systems can form the $JI$ states with $J=1/2,3/2$ and $I=0,1$.

In the following, we discuss the $ND^{(*)}$ and $N\bar{B}^{(*)}$ systems separately.

\subsection{$ND^{(*)}$ systems}

First we discuss the $ND^{(*)}$ systems ($ND$ and $ND^*$) in  the bound kinematics and then in the unbound (scattering) kinematics.

Since neither the OGE nor the confinement
potentials between the two color-singlet clusters $N$ and
$D^{(*)}$ are present, as mentioned in Sec. II, the effective potential between the two clusters arises only from
the one-meson-exchange interactions, $V^{ch}_{qq}$, between the light quarks.
The OGE and confinement potentials are present in the calculations of the individual clusters $N$ and  $D^{(*)}$ (in particular, their masses), however.
Figure \ref{vd12} displays the effective potentials between the $N$ and $D$, and, $N$ and $D^*$ clusters in the $S$-wave $[ND]_{J=1/2}$, $[ND^*]_{J=3/2}$ and $[ND^*]_{J=1/2}$ states, respectively,
as functions of $s$, the generator coordinate which describes, qualitatively, the distance between the two clusters in the generator coordinate method (GCM)
calculation\cite{o,p,q,r}.
In Fig.~\ref{vd12}, the dotted (red) curves correspond to the scalar $\sigma$-meson exchange contribution, while
the dashed (green) curves represent the sum of the scalar $a_0$- and $f_0$-meson contributions (no $\sigma$ meson contribution included here).
The dash-double-dotted (magenta) curves correspond to the sum of the pseudoscalar $\pi$-, $\eta$- and $\eta'$-meson and the dash-dotted (blue) curves to the sum of the vector $\rho$- and $\omega$-meson contributions.
Note that the $\phi$-meson contribution vanishes identically because, as mentioned in Sec.~\ref{sec:Formalism}A,  it is taken to be a pure $s\bar{s}$ state in the present work. The net contribution of all mesons is represented by the solid (black) curves.
We mention that the contribution of the pseudoscalar meson $\eta'$ is negligible compared to the other meson contributions, hence, in the following, we shall make no reference to this meson contribution.

As can be seen, the net (total) effective potential is attractive in all the cases. Except for the case of $[ND^*]_{J=1/2}$ with $I=1$, the bulk of the attraction is provided by the isoscalar $\sigma$-meson exchange. Compared to the chiral SU(3) quark model (set I), the net effective potential is more attractive in the extended chiral SU(3) quark model (sets II and III) due to the presence of the vector-meson ($\rho$ plus $\omega$) exchange contribution which provides an additional attraction. Note, in particular, that the vector-meson contribution is comparable to that of the $\sigma$-meson in the case of $[ND^*]_{J=1/2}$ with $I=1$; this is due to the isovector $\rho$-meson  which provides a much stronger attraction compared to the cases of  $[ND]_{J=1/2}$ and $[ND^*]_{J=3/2}$.  The pseudoscalar-meson exchange contribution is largest in the case of $[ND^*]_{J=1/2}$ with $I=1$ due to the $\pi$-meson providing the bulk of the repulsion, comparable in magnitude to those of the $\sigma$- and vector-mesons. It cancels partly the otherwise very strong attraction arising from the $\sigma$- and vector-meson contributions in the extended chiral SU(3) quark model. In the chiral SU(3) quark model, due to the absence of the attractive vector-meson potential, the resulting (attractive) net potential is much shallower than that in the former model.  As we shall show later, this has a direct consequence in the formation of bound states in the $ND^*$ system with quantum numbers $J=1/2$ and $I=1$.  Note that the pseudoscalar-meson exchanges are absent in the case of  $[ND]_{J=1/2}$ because of the antisymmetric spin wave function of the $D$ cluster.

The aspects of the effective potentials between the $N$ and $D^{(*)}$ clusters discussed above are the dominant features exhibited by these potentials. We now look at some of the more detailed aspects of the underlying dynamics exhibited by our effective potentials.
Our analysis indicates that, for the isospin $I=0$ states, the scalar-isoscalar meson $f_0$ contributes a weak attraction while the scalar-isovector meson $a_0$, a repulsion comparable in magnitude to the attractive $f_0$-meson potential. These result in nearly vanishing contribution of the sum of the scalar-meson exchanges excluding the $\sigma$-meson, as shown in Fig.~\ref{vd12} (dashed (green) lines). For the $I=1$ states, both $f_0$ and $a_0$ provide weak attractions, leading to slightly attractive potentials, also shown as dashed (green) curves in Fig.~\ref{vd12}.
 Note that the potential from an isovector meson contribution for the $I=1$ state differ by a factor of -3 compared to the corresponding contribution for the $I=0$ state.
The effective potentials arising from the pseudoscalar mesons exhibit quite differently behavior depending on the $JI$ quantum numbers as well as on the systems $ND$ or $ND^*$. First of all, as has been mentioned before, for the $[ND]_{J=1/2}$ state with $I=0,1$, their contributions vanish.
For the $[ND^*]_{J=3/2}$, $I=1$ state, their overall contribution is weakly attractive due to the attraction from the $\pi$ meson being a bit larger than the repulsion from the $\eta$ meson. Here, the relatively small attractive individual contributions arising from the pseudoscalar and scalar ($a_0$ plus $f_0$) mesons add up to a significant attraction.
For the corresponding $I=0$ state, both the $\eta$ and $\pi$ mesons contribute a repulsion, adding up to a still relatively small net repulsion.
The situation in the case of the $[ND^*]_{J=1/2}$ states is reversed from the case of $[ND^*]_{J=3/2}$ in the chiral SU(3) quark model(set I) as far as the total isospin $I$ dependence is concerned.
Furthermore, the contribution is enhanced by more than a factor of $2$ in the $I=1$ state as compared to the case of the $[ND^*]_{J=3/2}$, $I=0$ state.
Here, for the $I=0$ state, the overall contribution of the pseudoscalar mesons is attractive due to the attraction from $\pi$ being larger than the repulsion from $\eta$.
However, for the $I=1$, $\eta$ and $\pi$ contribute a relatively strong repulsion, comparable in magnitude to the strong attraction arising from the $\sigma$ meson.
In the extended chiral SU(3) quark model (sets II and III), we have the additional contributions to the effective potential from the vector mesons.  The $\omega$ meson contributes an attraction for all the states $[ND]_{J=1/2}$, $[ND^*]_{J=1/2}$ and $[ND^*]_{J=3/2}$, The $\rho$-meson contributes a weaker repulsion for the $[ND]_{J=1/2}$ and $[ND^*]_{J=1/2}$ states with $I=0$, while an attraction for the corresponding states with $I=1$. This leads to a net attractive vector-meson contribution for $I=0$ that is weaker than the corresponding attraction for $I=1$. Note, in addition, that the attraction for the $[ND^*]_{J=1/2}$, $I=1$ state is much larger than that for the $[ND]_{J=1/2}$, $I=1$ state. For $[ND^*]_{J=3/2}$, the vector mesons contribute a net attractive potential for both $I=0$ and $I=1$. However, the attraction for $I=1$ is weaker than for $I=0$, which is opposite to what is observed for the corresponding states in $[ND^*]_{J=1/2}$ as far as the isospin dependence is concerned. Finally, we point out that the relatively small difference in the corresponding vector-meson contributions between set II and set III, is sorely due to the $\omega$-meson tensor coupling as shown in Table.~\ref{para}.

The features of the effective potentials exhibited in Fig.~\ref{vd12} and discussed above are reflected in the calculated binding energies of the possible bound states in the $S$-wave $ND$ and  $ND^*$ systems displayed in Table.\ref{bind}, corresponding to the three sets of model parameters given in Table.~\ref{para}.

\begin{table}[htb]
\caption{{\label{bind}}Binding energies $E_b$(MeV) of possible $S$-wave $ND^{(*)}$. The last column is the results from Ref.\cite{pgo-16}.}
\setlength{\tabcolsep}{1.0mm}
\begin {center}
\centering
\begin{tabular}{ccccccc}
\hline\hline&&& $\chi$-SU(3)QM& \multicolumn{2}{c}{Ex. $\chi$-SU(3) QM}& \\
& $J^P$&Isospin& I & II& III& Ref.\cite{pgo-16}\\
 \hline \centering
 $ND$      &$\frac{1}{2}^-$ &0&$-$ &10.0&5.9& 1.70  \\
           &&1&2.2 &38.8 &26.7&\\

 $ND^*$     &$\frac{1}{2}^-$  &0&2.0 &18.1 &12.8& \\
           &&1&$-$&27.5 &17.6& 0.48 \\

      &$\frac{3}{2}^-$  &0&$-$ &6.6 &3.2&8.02 \\
           &&1&8.0 &44.8 &31.6&\\
 \hline\hline
\end{tabular}
\end{center}
\end{table}

The binding energy $E_b$ of the
nucleon-meson ($NM_Q$) system is defined as
\begin{equation}
E_{b}=-\left[M_{NM_Q}-(M_N+M_{M_Q})\right].
\label{eq:BE}
\end{equation}
$M_{NM_Q}$, $M_N$, and $M_{M_Q}$ are the calculated masses of the five-quark system $[NM_Q]$, nucleon $N$, and heavy meson $M_Q$, respectively.
If $E_b$ is positive, the system is bound as a molecular state.

Table \ref{bind} displays all the bound states predicted in the present work with the corresponding binding energies.
As can be seen, in the extended chiral SU(3) quark model(set II and III), we obtain two bound states in the $[ND]$ system, namely, $[ND]_{J=1/2}$ with $I=0$ and $I=1$, and four bound states in the $ND^*$ system, two with spin-parity $J^P=1/2^-$ and two with $J^P=3/2^-$, i.e., $[ND^*]_{J=1/2}$ with $I=0$ and $I=1$, and $[ND^*]_{J=3/2}$ with $I=0$ and $I=1$.  In contrast, in the chiral SU(3) quark model (set I), no bound states $[ND]_{J=1/2}$ and $[ND^*]_{J=3/2}$ with $I=0$ and  $[ND^*]_{J=1/2}$ with $I=1$ are found. This is a direct consequence of the absence of the attractive vector-meson exchange potential  in the later model, which makes the resulting total effective potentials much shallower than in the extended SU(3) quark model, as can be seen in Fig.~\ref{vd12}. Note, in particular, that the binding energies of the predicted states in the chiral SU(3) quark model are much smaller than the corresponding binding energies predicted in the extended chiral SU(3) quark model. Furthermore, the binding energies corresponding to set II are larger than those of set III, in accordance with the stronger attractive potential in set II than in set III as can also be seen in Fig.~\ref{vd12}.

We know that the experimentally observed $\Sigma_c(2800)$ is about $8$ MeV below the $ND$ threshold.
Therefore, the predicted $[ND]_{J=1/2}$ with $I=0$ molecular state with $10.0$ MeV(set II) or $5.9$ MeV(set III) binding energy in the extended chiral SU(3) quark model is a likely candidate for the $\Sigma_c(2800)$ state. Its spin-parity $J^P=1/2^-$ is
consistent with the findings from other independent calculations \cite{cej-09,jrz-14,zyw-18},
as well as with the weak evidence of $J=1/2$ from Babar Collaboration\cite{babar-08}.
On the other hand, the isospin of our $\Sigma_c(2800)$ candidate is $I=0$, which is at odds with the experimentally inferred value of $I=1$\cite{guo-18,belle-05}.
We expect further experiments to confirm or dismiss our results.

Similarly, the experimentally observed $\Lambda_c(2940)^+$ is about $6$ MeV below the  $ND^*$ threshold. Both the present chiral and extended chiral SU(3) quark models predict an $ND^*$ bound state with $J^P=3/2^-$ in this energy region. The former model yields an $[ND^*]_{J^P=3/2^-}$ state with $I=1$ at about 8.0 MeV (set I) binding energy, while the latter model gives an $[ND^*]_{J^P=3/2^-}$ state with $I=0$ at about 6.6 MeV (set II) or 3.2 MeV (set III) binding energy. The recent experiment \cite{lhc-17} assigns the spin-parity $J^P=3/2^-$  while the  other experiments \cite{guo-18,babar-07,belle-07,pdg} assign the isospin $I=0$ to the $\Lambda_c(2940)^+$. This rules out the chiral SU(3) quark model, indicating that the extended chiral SU(3) quark model is more effective in predicting the $S$-wave bound states in the $ND^*$ system, which means that the vector-meson exchange interactions play a significant role.

At this point, we mention that the authors of Ref.\cite{pgo-16} have also investigated the possible $ND$ and $ND^*$ molecules in a chiral constituent quark model, and their results are listed in the last column of Table \ref{bind}. They suggest the $[ND^*]_{J=3/2}$ with $I=0$ with $8.02$ MeV binding energy as the observed $\Lambda_c(2940)^+$, and don't find the candidate for the $\Sigma_c(2800)$. We corroborate these findings. Our binding energies for the $[ND]_{J=1/2}$ with $I=0$ and $[ND^*]_{J=1/2}$ with $I=1$ states are larger than theirs.
In contrast to the present work, where we have included explicitly the low-lying vector-meson nonet into the calculation (extended chiral SU(3) quark model), in Ref.~\cite{pgo-16}, the vector-mesons are not considered based on the argument to avoid double counting because the vector mesons provide the short-range interaction, which is taken over by the OGE potential \cite{ava-05}. As we have pointed out, however, the OGE potential does not act between the $N$ and $D^{(*)}$ or $\bar{B}^{(*)}$ clusters.
Note that the possible double-counting issue that arises in the light quark flavor sector, where both the OGE and vector-meson exchange contribute is avoided to the extent that the OGE coupling is reduced when the vector mesons are also included to always reproduce the relevant physical quantities (See the value of the coupling constant $g_u$ in Table.~\ref{para} which is reduced by nearly a factor of 4 in going from the chiral to extended chiral SU(3) quark model.)

We now turn our attention to the issue of the flavor single-octet mixing of the scalar mesons $\sigma_0$ and $\sigma_8$. As mentioned in Sec.~\ref{sec:Formalism}.A, both the chiral and extended chiral SU(3) quark models applied in the present work ignore the $\sigma_0$-$\sigma_8$ mixing, i.e., $\theta^s= 0^{\circ}$.  These models were built to describe successfully many hadronic systems in the light flavor sector \cite{xcqm-zhang, lrdai03, zd,ls}.
 The present analysis of the $ND^{(*)}$ systems reveals that  the $\sigma$ meson has a very important contribution to the effective potential between the $N$ and $D^{(*)}$ cluster, providing the necessary attractions to form bound states (cf.~Fig.~\ref{vd12}). It happens that the resulting attraction in the effective potential and, consequently, the binding energies are rather sensitive to the mixing angle $\theta^s$.
Unfortunately,  the mixing angle $\theta^s$ is not as well established as the corresponding angles for the pseudoscalar and vector mesons. The analysis of Ref.~\cite{jao-03}, based on Unitarized Chiral Perturbation Theory, have shown that singlet-octet mixing holds in the scalar-meson sector and quotes a value of $\theta^s=19^{\circ} \pm 5^{\circ}$. The positive sign of $\theta^s$ is preferred over the negative sign, since the latter sign-choice leads to a large U(3) symmetry breaking. On the other hand, in Ref.~\cite{ybd-04}, based on an approximate chiral symmetry effective Lagrangian, the mixing angle of $\theta^s=-18^{\circ}$ is quoted.  These very distinct values of $\theta^s$ reveal that the issue of $\sigma_0$-$\sigma_8$ mixing is far from being settled.
We emphasize that a proper assessment of the sensitivity of the present results to $\theta^s$ should be performed in a way that all the other parameter values of the model are consistent with the chosen value of $\theta^s$. This requires refitting those parameters to describe the relevant hadronic systems in the light flavor sector which is beyond the scope of the present paper.  We reserve this for a future work.
In this regard, it is noteworthy that, in a recent chiral SU(3) quark model study by Huang and Wang\cite{hf18},
the masses of the octet and decuplet baryon ground
states, the binding energy of deuteron, and the $NN$
scattering phase shifts have been simultaneously reproduced in a rather consistent manner.
Specifically, the harmonic-oscillator size parameters for constituent quarks were determined
by a variational method instead of being treated as predetermined parameters and taken to be
the same for all single baryons. This ensures that all single
baryons are minima of the Hamiltonian.
It would be interesting to consider their model as a basis for including the $ND^{(*)}$ and $N\bar B^{(*)}$ systems.

\begin{figure}[htb]
\centerline{\includegraphics[height=11cm,width=8.5cm]{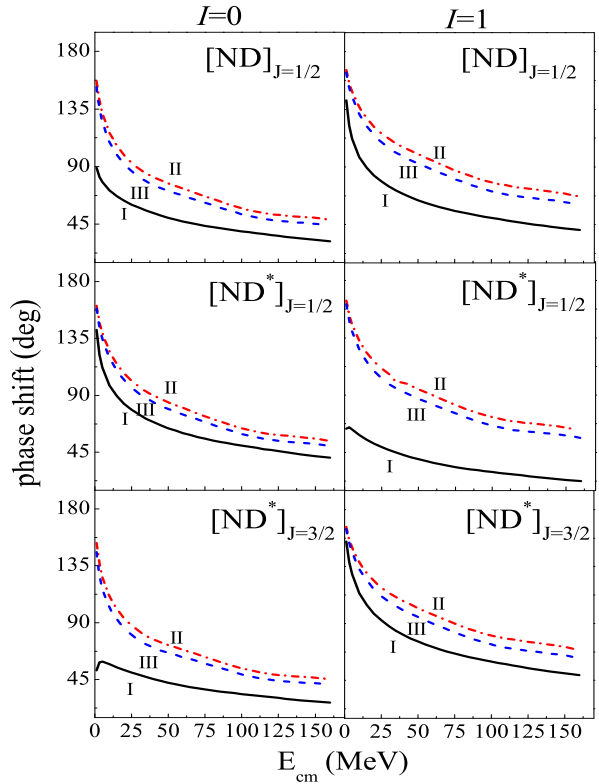}}
\caption{\small \label{phase} $ND^{(*)}$ $S$-wave phase-shifts
as functions of the center-of-mass energy. The solid lines
represent the results in the chiral SU(3) quark model (set I),
and the dash-dotted and dashed lines represent the results in the
extended chiral SU(3) quark model (sets II and III). Panels on the left are for $I=0$ and those on the right are for $I=1$.}
\end{figure}

In the present work, we also study the $S$ partial-wave $ND^{(*)}$ elastic scattering processes
by solving the RGM equation in order to obtain information on the underlying $ND^{(*)}$ interactions in the scattering kinematics.
The calculated phase-shifts are illustrated in Fig.\ref{phase} as a function of c.m. energy $E_{\rm{cm}}$.
As can be seen in Fig.\ref{phase}, the $S$-wave phase-shifts corresponding to the parameter sets I(chiral SU(3) quark model), II, and III(extended chiral quark model) are all positive, which means that the corresponding underlying interactions are all attractive. They decrease rapidly as functions of energy.
The magnitudes of the $S$-wave phase-shifts in the extended chiral SU(3) quark model are larger than that of the chiral SU(3) quark model, especially that corresponding to the parameter set II.
Also, magnitudes for $I=1$ are larger than the corresponding ones with $I=0$, except for the case of $[ND^*]_{J=1/2}$ corresponding to set I.
The larger is the phase-shift, the greater is the attraction.

\subsection{$N\bar{B}^{(*)}$ systems}

The analysis of the $ND^{(*)}$ systems in the preceding subsection can be carried over to the $N\bar{B}^{(*)}$ systems, by replacing the $c$ quark by the $b$ quark.
The behaviors of the resulting effective interactions between $N$ and $\bar{B}^{(*)}$ are similar to those exhibited by the $ND^{(*)}$ interactions shown in Fig.~\ref{vd12}, except for the fact that the attractions between $N$ and $\bar B^{(*)}$ are a bit larger than those between $N$ and $D^{(*)}$.
The binding energies of all possible $S$-wave $N\bar{B}^{(*)}$ states obtained are tabulated in Table \ref{bind2}.
The $[N\bar B]_{J=1/2}$ states with $I=0$ and $I=1$ are both bound with the binding energies in the range of $1-53$ MeV depending on the models and parameter sets (I, II, and III) considered.
The $[N\bar B^*]_{J=3/2}$ bound state with $I=0$ is found only in the extended chiral SU(3) quark model
with binding energy of $13.7$ MeV and $8.9$ MeV, corresponding to the parameter sets II and III, respectively.
On the other hand, the $[N\bar B^*]_{J=3/2}$ bound state with $I=1$ is found in both the chiral and extended chiral SU(3) quark models with the binding energies in the range of $16-60$ MeV, depending on the models and parameter sets.
Similarly, the $[N\bar B^*]_{J=1/2}$ states with $I=0$ and $I=1$ are bound with the binding energy in the range of $7-40$ MeV. The exception is the bound $[N\bar B^*]_{J=1/2}$ state with  $I=1$ which is absent in the chiral SU(3) quark model (set I).
One sees that the binding energies of the $N\bar{B}^{(*)}$ states are larger than those of the corresponding  $ND^{(*)}$ states.

Also, as shown in Table \ref{bind2}, Ref.\cite{pgo-16} found four $N\bar B^{(*)}$ bound states. Our binding energies of $[N\bar B]_{J=1/2}$ and $[N\bar B^*]_{J=3/2}$ with $I=0$ are close to their corresponding results, while those of $[N\bar B]_{J=1/2}$ with $I=1$ and $[N\bar B^*]_{J=1/2}$ with $I=0$ are larger than theirs.

\begin{table}[htb]
\caption{{\label{bind2}}Binding energies $E_b$(MeV) of possible $S$-wave $N\bar{B}^{(*)}$(MeV). The last column is the results from Ref.\cite{pgo-16}.} \setlength{\tabcolsep}{1.0mm}
\begin {center}
\centering
\begin{tabular}{ccccccc}
\hline\hline&&& $\chi$-SU(3)QM& \multicolumn{2}{c}{Ex. $\chi$-SU(3) QM}&\\
& $J^P$&Isospin& I & II& III& Ref.\cite{pgo-16}\\
 \hline \centering
 $N\bar{B}$      &$\frac{1}{2}^-$ &0&1.1 &18.2&12.8&12.09  \\
           &&1&7.8 &53.2 &39.3&0.36\\

 $N\bar{B}^*$     &$\frac{1}{2}^-$  &0&7.7 &28.6 &22.0&3.43 \\
           &&1&$-$&40.3 &28.5& \\

      &$\frac{3}{2}^-$  &0&$-$ &13.7 &8.9&15.15\\
           &&1&16.3 &59.9 &44.9&\\
 \hline\hline
\end{tabular}
\end{center}
\end{table}

The results of scattering process of $N\bar{B}^{(*)}$ and $ND^{(*)}$ are quite alike.
The $S$-wave phase-shifts in $N\bar{B}^{(*)}$ scattering have similar behavior to those corresponding ones in $ND^{(*)}$ shown in Fig.\ref{phase} with
slightly larger magnitudes. The difference is no more than $10$ degrees. This feature indicates that the attraction between $N$ and $\bar B^{(*)}$ is greater than that between $N$ and $D^{(*)}$, which is also the feature seen in the binding energies in Tables.\ref{bind} and \ref{bind2} in the bound state kinematics.

\section{Summary and conclusion}

We have explored some of the properties of the
$ND^{(*)}$ and $N\bar{B}^{(*)}$ systems  in the $S$ partial waves by solving the RGM equation in both the chiral and extended chiral SU(3) quark models,
including the bound-state and elastic scattering processes.
We have found that the effective potential between the two clusters, $N$ and $D^{(*)}$ or
$N$ and $\bar{B}^{(*)}$, is attractive, and this
attraction in the extended chiral SU(3) quark model (specially in
set II) is stronger than that in the chiral SU(3) quark model.
Also,  the attraction is stronger for the $I=1$ states than for the $I=0$ states, except for the case of the
$[ND^*]_{J=1/2}$ and $[N\bar B^*]_{J=1/2}$ states in the chiral SU(3) quark model, where the attraction is stronger for the $I=0$ states than for the $I=1$ states.
The attractive nature of the effective potential has been shown to arise primarily from the $\sigma$-meson exchange in both the chiral and extended chiral SU(3) quark models for $[ND]_{J=1/2}$, $[ND^*]_{J=3/2}$ with both $I=0,1$ and $[ND^*]_{J=1/2}$ with $I=0$. For $[ND^*]_{J=1/2}$ with $I=1$ there is a sizable repulsive contribution from the pion exchange and, in the extended chiral SU(3) quark model, an attractive vector meson contribution comparable to that of the $\sigma$-meson.
In the later model, the vector meson ($\rho$, $\omega$) exchanges provide further attraction.
This extra attraction from the vector mesons suffices to form the bound states $[ND]_{J=1/2}$ with $I=0$,
$[ND^*]_{J=1/2}$ with $I=1$ and $[ND^*]_{J=3/2}$ with $I=0$, which are absent in the chiral SU(3) quark model.
Analogously, the bound states $[N\bar{B}^*]_{J=1/2}$ with $I=1$ and $[N\bar{B}^*]_{J=3/2}$ with $I=0$ are only formed in the extended chiral SU(3) quark model.

According to the present model analysis, the observed $\Sigma_c(2800)$ and $\Lambda_c(2940)^+$ may be interpreted, respectively, as the $S$-wave $ND$ molecular state with $I=0$, $J^P=1/2^-$ and the $S$-wave $ND^*$ molecular state with $I=0$, $J^P=3/2^-$. This finding indicates the extended chiral SU(3) quark model is more effective in describing the $S$-wave $ND^{(*)}$ system as compared to the chiral SU(3) quark model. Although more in depth studies are required, we would like to point out that the basic qualitative features of all possible $S$-wave $ND^{(*)}$ and $N\bar{B}^{(*)}$ states obtained in this study are reasonable when compared to the currently available experimental and theoretical information. Also, the future experimental search for the $N\bar{B}^{(*)}$ molecular states is an interesting topic.

The calculated $S$-wave phase-shifts corresponding to the
$ND^{(*)}$ and $N\bar{B}^{(*)}$  scattering processes reveal that the corresponding interactions are also quite  attractive , decreasing rapidly as the energy increases.

Finally, we mention that we have found that the $S$-wave bound states in the $ND^{(*)}$ and $N\bar{B}^{(*)}$ systems are sensitive to the flavor singlet-octet mixing angle of the scalar meson nonet. Unfortunately the corresponding mixing angle $\theta^s$ is not well known. This imposes a limitation on the predictive power of the present type model, where this parameter value is required. A proper assessment of the sensitivity of the calculated binding energies to $\theta^s$ requires a re-analysis of the relevant hadronic systems in the light flavor sector to consistently determine all the parameters of the model as a function of $\theta^s$. This is beyond the scope of the present work and we reserve such analysis to future study. In this connection, models where the $\sigma$ meson is calculated as the $\pi\pi$ system in the $J=I=0$ state could avoid the problem associated with the mixing angle $\theta^s$.

\begin{acknowledgements}
This work was supported by the Natural Science Foundation of Inner
Mongolia Autonomous Region of China (2015MS0115). D.Z. thanks Dr. Fei Huang for helpful discussions
and for a careful reading of the manuscript. D.Z. also acknowledges
the kind hospitality of the Department of Physics and Astronomy at the University of Georgia during her visit, where the part of this work has been carried out.
\end{acknowledgements}

\end{document}